\documentclass[amssymb,aps,twocolumn,floats,showpacs]{revtex4-1}
\usepackage{color,graphicx,pstricks,float}
\usepackage{natbib}

\newcommand{\Fig}{Fig~}

\begin{document}
\title{Coupled spin-charge order in frustrated itinerant triangular magnets}

\author{Sahinur Reja$^1$, Rajyavardhan Ray$^2$, Jeroen van den Brink$^1$ and Sanjeev Kumar$^2$}
\email{sanjeev@iisermohali.ac.in}
\address{
$^1$ Institute for Theoretical Solid State Physics, IFW Dresden, 01171 Dresden, Germany \\
$^2$ Indian Institute of Science Education and Research (IISER) Mohali, Sector 81, S.A.S. Nagar, Manauli PO 140306, India \\
}

\begin{abstract}
We uncover four new spin-charge ordered ground states in the strong coupling limit of the Kondo 
lattice model on triangular geometry.
Two of the
states at one-third electronic filling ($n=1/3$) consist of decorated ferromagnetic chains 
coupled antiferromagnetically with the neighboring chains. 
The third magnetic ground state is noncollinear, consisting of antiferromagnetic chains separated by 
a pair of canted
ferromagnetic chains. An even more unusual magnetic ground state, a variant of the $120^{\circ}$ Yafet-Kittel phase, is discovered at $n=2/3$.
These magnetic orders are stabilized by opening
a gap in the electronic spectrum: a ``band effect''.
All the phases support modulations in the electronic charge density due to the presence of magnetically inequivalent sites. 
In particular, the charge ordering pattern found
at $n=2/3$ is observed in various triangular lattice systems, such as, 2H-AgNiO$_2$, 3R-AgNiO$_2$ and Na$_x$CoO$_2$.
\end{abstract}

\date{\today} 


\maketitle

The influence of conduction electrons on the behavior of a system of localized magnetic moments is a 
well-studied topic in solid state physics. Investigations of such spin-charge coupled systems have given rise to 
a number of key concepts in magnetism and transport, such as, the RKKY interactions, the Kondo effect and
the double-exchange (DE) mechanism \cite{Ruderman1954, Kasuya1956, Yosida1957, Kondo1964, Zener1951, DeGennes1960}. 
These concepts are commonly invoked in order to understand magnetism and charge transport in 
materials ranging from dilute magnetic semiconductors to various transition metal oxides and heavy fermion compounds \cite{Dietl2014,Stewart1984}. 
In recent years, it has been realized that the geometry of the underlying lattice
plays a crucial role in determining the nature of magnetic states in such systems \cite{Ishizuka2012,Ishizuka2013,Martin2008,Chern2010}. 
In particular, geometrically frustrated lattices support unusual non-collinear and 
even non-coplanar spin-textures in the ground states \cite{Martin2008, Akagi2010a, Motome2010, Kumar2010, Chern2010, Venderbos2012}. 
The electronic response is dramatically affected by these
unusual spin-textures, exhibiting remarkable phenomena such as colossal magnetoresistance, anomalous and quantum 
anomalous Hall effects, and multiferroicity \cite{Salamon2001,Nagaosa2010,Cheong2007}. 
As a result of such diversity of phenomena associated with
unusual spin textures, their search in models, materials and artificial structures
has become a very active field of research \cite{Banerjee2013, Wang2014, Liu2014, Wang2006, Nisoli2013}.

The starting point for a theoretical analysis of the interplay between spin-charge coupling and magnetic frustrations is the 
Kondo-lattice model (KLM) on various frustrated geometries. In the limit of weak Kondo coupling, the shape of Fermi surface can play a crucial role
in determining the magnetic ground state \cite{Martin2008,Hayami2014}. Moreover, a perturbative expansion of free energy to various orders in 
Kondo-coupling can be used to derive
effective magnetic Hamiltonians \cite{Akagi2012}. However, in strong coupling limit, the relevance of a non-interacting Fermi surface or
that of a perturbative effective Hamiltonian in determining magnetic ground states is less clear. Nevertheless, there are many 
examples where the magnetic order in the strong coupling limit turns out to be the same as that in the weak coupling 
limit \cite{Martin2008, Akagi2010a, Kumar2010}.

The focus of this letter is the strong coupling limit of the KLM on triangular lattice.
We establish the presence of four exotic spin-charge ordered ground states at filling fractions of $n=1/3$ and $n=2/3$.
Two of these phases are collinear, and consist of decorated ferromagnetic chains. The other two phases are noncollinear (NC), of which one can
be visualized as
AFM chains separated by a pair of canted-FM chains. The other NC phase is similar to $120^{\circ}$ state, except that it consists of 
three type of spin triangles. An inequivalence between the lattice sites is induced by the peculiar spin ordering, 
causing an ordering of the electronic charge density. While the charge modulations are weak for phases at $n=1/3$, a
strong charge ordering is found at $n=2/3$ with an ordering pattern similar to that observed in experiments on various 
triangular lattice systems \cite{Wawrzynska2008,Wawrzynska2007a,Bernhard2004,Chung2008}.
All the magnetic phases are insulating with a gap of the order of hopping parameter. 
We show that two of the four phases are further stabilized by Coulomb repulsions. 
The existence of such novel spin-charge orderings
in a realistic model could guide the experimental search for unusual magnetic ordering phenomena.

We start from a strong coupling KLM, which reduces to a double exchange (DE) model with antiferromagnetic (AFM) exchange
between the localized moments. Assuming the local moments to be classical, we directly write down the 
Hamiltonian on the triangular lattice as,

\begin{eqnarray}
H &=& -\sum_{ \langle ij \rangle}
t_{ij} \left ( c^{\dagger}_{i} c^{~}_{j} + H.c. \right ) 
+ J_{AF} \sum_{ \langle ij \rangle} {\bf S}_{i} \cdot {\bf S}_{j},
\end{eqnarray}

\noindent
where $c^{}_{i}$ ($c^{\dagger}_{i}$) is the usual annihilation (creation) 
operator for electron with spin parallel to the local magnetic moment ${\bf S}_i$. 
The angular brackets in the summations denote the nearest neighbor (nn) pairs of sites on a triangular lattice. 
$J_{AF}$ is the strength of AF coupling between nn localized spins.
Note that $t_{ij}$ depend on the polar and azimuthal angles 
\{$\theta_i$,$\phi_i$,$\theta_j$,$\phi_j$\} of the nn
core spins, and are given by
$t_{ij} = t_0 [\cos(\theta_i/2)\cos(\theta_j/2) + \sin(\theta_i/2)\sin(\theta_j/2) e^{{\rm -i} (\phi_i - \phi_j)}]$.
The parameters of the model are, the hopping amplitude $t_0$,
the AF coupling $J_{AF}$ and the electronic filling fraction $n$. We set $t_0 = 1$ as the reference energy scale.

The model is investigated using the state of the art Monte Carlo (MC) method which combines
the classical MC for spins with numerical diagonalization for fermions \cite{Yunoki1998}.
The solution of a fermionic
problem is carried out numerically at each MC update step in order to obtain the 
electronic contribution to the total energy of a given classical spin configuration.
We have used $6 \times 6$ and $12 \times 12$ clusters for this study, with 
typically $10^4$ MC steps for equilibration and an equal number of steps for averaging.
The results obtained on these small clusters provide us with important clues about the 
nature of magnetic states. The energies of the candidate states are 
computed on much larger lattices in order to rule out any finite-size effects. 


 
The important physical quantity that contains information about the nature of magnetic ordering
is the spin structure factor ($S({\bf q})$), which is defined as,
\begin{eqnarray}
S({\bf q}) &=& \frac{1}{N^2} \sum_{ij} \langle {\bf S}_i \cdot {\bf S}_j \rangle_{av} ~ 
e^{-{\rm i}{\bf q}\cdot({\bf r}_i - {\bf r}_j)}. 
\end{eqnarray}

\begin{figure}[t!]
\centerline{
\includegraphics[width =.96\columnwidth , clip=true,angle=0]{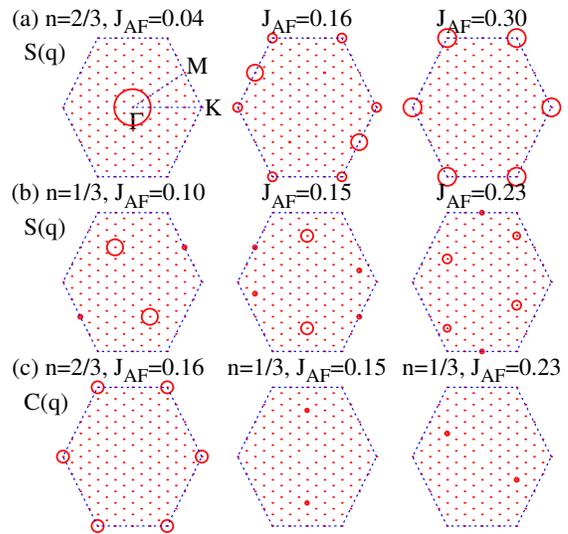}
}
\caption{(Color online) (a)-(b) Low-temperature spin structure factor for different values of $J_{AF}$
at $n=2/3$ and $n=1/3$. (c) Charge structure factor for three representative values of $J_{AF}$ at $n=2/3$ and $n=1/3$.
The circle size at a given ${\bf q}$ represents the magnitude of the structure factor at that ${\bf q}$.
}
\label{spin_SF}
\end{figure}

\noindent
In the above $\langle ... \rangle_{av}$ denotes the thermal or MC average, $N$ is the number of lattice sites, and 
${\bf r}_i$, ${\bf r}_j$ are the position vectors of sites $i$, $j$.
We begin by discussing the spin structure factor results
obtained from simulations on $12 \times 12$ lattice. 
\Fig\ref{spin_SF} (a) and (b) display the results at low temperature, $T = 0.002 t_0$,
for filling fractions $n=2/3$ and $n=1/3$ respectively. The magnitude of $S({\bf q})$ is indicated by the radius of the empty circles, and
the ${\bf q}$ values are restricted to the 1st Brillioun zone.
In the low-$J_{AF}$ limit, the $S({\bf q})$
peaks at the $\Gamma$ point indicating a ferromagnetic ground state, which is expected in the DE model. For $n=1/3$, 
the peak at the $\Gamma$ point
remains robust in the range $0 \leq J_{AF} < 0.08$. For $J_{AF} = 0.10$, we find two peaks in the $S({\bf q})$ 
(see \Fig\ref{spin_SF} (b)), one at the 
$M$-point and other on the $\Gamma$-$K$ axis. 
For $J_{AF} = 0.16$, $S({\bf q})$ indicates the presence of another unusual magnetic phase with 
peaks at multiple ${\bf q}$ points.
We confirm by looking at the spin configurations in real space that both these phases are collinear.
The $S({\bf q})$ at $J_{AF} = 0.23$ is qualitatively different, indicating the appearance of yet another magnetic order.
We will discuss the nature of these phases in detail later. At this point, we emphasize these magnetic phases
have not been reported in any of the previous studies on Kondo lattice model on triangular lattice.
Also, the plots at different values of $J_{AF}$ 
are only representative of different phases. The stability range of these phases will become clear when we discuss
the phase diagrams.
 
For $n=2/3$, the
presence of another unusual magnetic order in the coupling range $0.04 < J_{AF} < 0.20$ is inferred 
from the $S({\bf q})$. This phase is characterized by
two peaks in the $S({\bf q})$ at the $K$ and $M$ points (see \Fig\ref{spin_SF} (a) plotted for $J_{AF} = 0.16$).
It is also clear from the structure factor plots that all the new phases discussed above break the three-fold rotational symmetry of the
triangular lattice.
In order to further probe the nature of electronic states in these new magnetic phases, 
we compute charge structure factor defined as,
\begin{eqnarray}
C({\bf q}) &=& \frac{1}{N^2} \sum_{ij} \langle \delta n_i \delta n_j \rangle_{av} ~ 
e^{-{\rm i}{\bf q}\cdot({\bf r}_i - {\bf r}_j)},
\end{eqnarray}
\noindent 
where $\delta n_i = n_i - n$ is the charge density modulation w.r.t. the average charge density $n$.
The $C({\bf q})$ plots in \Fig\ref{spin_SF} (c) show that all the magnetic phases discussed above 
exhibit charge ordering. For the phases at $n=1/3$, the magnitude of
charge disproportionation is small, and the ordering pattern is stripe-like. However, for the 
NC magnetic phase at $n=2/3$, the charge ordering is strong in magnitude, and has a pattern similar to the one
observed in various triangular lattice systems with active spin degree of freedom, such as,
2H-AgNiO$_2$, 3R-AgNiO$_2$ and Na$_x$CoO$_2$ \cite{Wawrzynska2008,Wawrzynska2007a,Bernhard2004,Chung2008}.
Typically a CO state arises
either due to Coulomb repulsions at appropriate filling fractions, or due to charge-lattice couplings \cite{kumar2008charge}.
Therefore, it is unusual that charge ordering emerges in a model consisting of local charge-spin coupling. Indeed,
this was emphasized in a recent work reporting the presence of 
an unusual spin-charge ordered state in KLM \cite{Misawa2013}.


\begin{figure}[t!]
\centerline{
\includegraphics[width =.96\columnwidth , clip=true,angle=0]{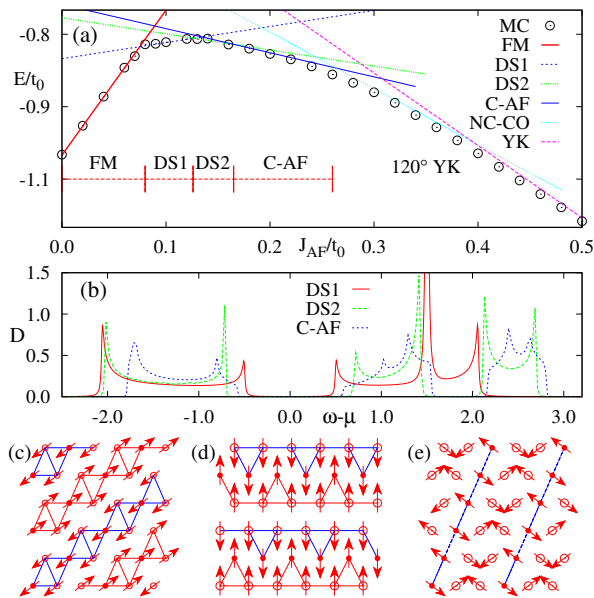}
}
\caption{(Color online) (a) Energy per site at $T/t_0 = 0.002$ as a function of $J_{AF}$ obtained via MC simulations (circles) for a 
filling fraction of $n=1/3$. Various straight lines are the energies of different phases as indicated by legends. 
(b) The electronic density of states for three new ground states, DS1, DS2, and C-AF.
(c)-(e) Snap-shots of the MC configurations for the three new ground states.
Arrows indicate the spin directions, and the circle sizes indicate the local charge density. The smaller circles have been filled to
highlight the pattern of charge ordering.
}

\label{diff_phases13}
\end{figure}

We now discuss in detail the phase diagram of the model at $n=1/3$. In Fig. ~\ref{diff_phases13} (a),
we plot the ground state energy for different $J_{AF}$ values obtained from MC simulations.
Looking at the low-T spin configurations from the simulations, we infer the nature of magnetic ground states for 
different $J_{AF}$. The straight lines in Fig. ~\ref{diff_phases13} (a) correspond to the energy obtained for ideal 
long-range ordered spin arrangements. 
For small $J_{AF}$, the MC energies fall on the straight line corresponding to a FM phase, as expected. Similarly, in the limit of 
large $J_{AF}$ the MC energies match well with those of the $120^{\circ}$-YK phase. We require four additional magnetic phases in order
to fit the MC data in the intermediate $J_{AF}$ regime. The spin arrangement for three of these magnetic phases are shown in Fig. ~\ref{diff_phases13} (c)-(e).
We label two of these phases as decorated stripes (DS1 and DS2), and the third as canted AF (C-AF). The fourth phase will be discussed later, as it turns out to be
the dominant phase at $n=2/3$.

In order to visualize these phases more clearly, we have connected all the ferromagnetically oriented spins via solid lines. This highlights the main feature of
DS1 (see Fig. ~\ref{diff_phases13} (c)), that this phase consists of diamond shaped FM chains running along one direction connected antiferromagnetically 
to the neighboring spins. Similarly, the DS2 phase consists of FM stripes decorated by triangular units (see Fig. ~\ref{diff_phases13} (d)).
In the strong Kondo coupling limit, the electronic hopping across a pair of sites hosting antiferromagnetically oriented spins is zero.
Therefore, in the DS1 and DS2 phases, the electronic problem becomes one-dimensional. 
The electronic density of states (DOS) in the decorated stripe phases has large gaps at the 
chemical potential (see Fig. ~\ref{diff_phases13} (b)). Opening of
these gaps in the DOS lowers the total energy of the system and hence these unusual phases are stabilized.
Such decorated stripe paths for hopping are realized in certain organic polymers \cite{VanderHorst1999,Cote2001}, and 
are also of interest to researchers working on exactly solvable models of electroic correlations
\cite{Gulacsi2007, Gulacsi2010}. Interestingly, such novel structures 
for fermion hopping can emerge in a higher dimensional lattice via a subtle interplay between geometrical frustrations and
spin-charge coupling. 

\begin{figure}[t!]
\centerline{
\includegraphics[width =.96\columnwidth , clip=true]{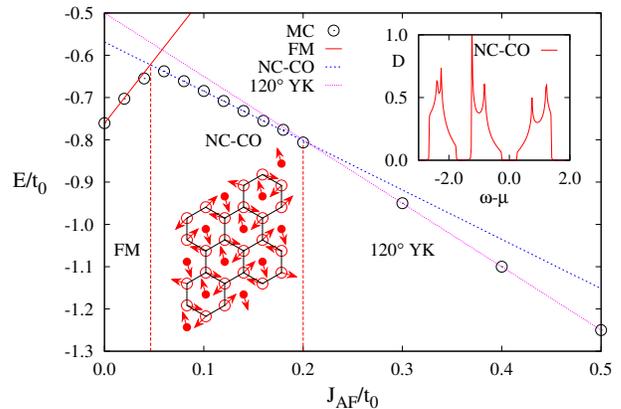}
}
\caption{(Color online) Groundstate energy per site as a function of $J_{AF}$ obtained via MC simulations (circles) for
filling fraction of $n=2/3$. The solid, dashed and dotted straight lines are the energies of ferromagnetic, NC-CO and the 
$120^{\circ}$ Yafet-Kittel states, respectively. The lower inset shows the snap-shot of the MC ground state with the arrows
representing the spin directions and the circle sizes indicating the local charge density. The smaller circles have been filled
and high charge density points are connected by lines to
highlight the real-space pattern of charge ordering. The inset in top-right corner shows the density of states for the NC-CO state.
}
\label{phase_diag23}
\end{figure}

The two phases discussed so far are collinear in nature and therefore allow for a description in terms of FM chains. The
third new phase at $n=1/3$ is NC, and consists of AF chains separated by a pair of canted FM chains. This spin arrangement also
opens a gap in the electronic DOS at the chemical potential. 
All the phases discussed above contain 
inequivalent sites in terms of the orientation of neighboring spins. 
This causes a modulation in the local charge density, and indeed we find a
charge-ordering in all the phases (see Fig. ~\ref{diff_phases13} (c)-(e)). 
For opening a gap in the spectrum, that we find in all the three phases discussed above, 
the entire magnetic structure must be modified. This can be seen as a ``band effect''. 
It is also interesting to note that the DS1 and DS2 spontaneously break a discrete rotational symmetry. 
Since there cannot be any magnetic order at finite temperatures in
a rotationally invariant continuous spin model \cite{Loss2011}, it may give rise to an interesting nematic magnetic order at finite
temperatures.
These new magnetic phases exist in the wide parameter range $0.08 < J_{AF} < 0.26$. For larger values of $J_{AF}$ the magnetic order gradually changes towards the 
120$^{\circ}$ Yafet-Kittel (YK) phase. We note that in the range $0.26 < J_{AF} < 0.36$ the energy of another unusual spin-charge ordered state
is close to that obtained from MC simulations.
In fact, the same state dominates the phase diagram at $n=2/3$ which we discuss next.


\begin{figure}[t]
\centerline{
\includegraphics[width =.98\columnwidth , clip=true, angle=0]{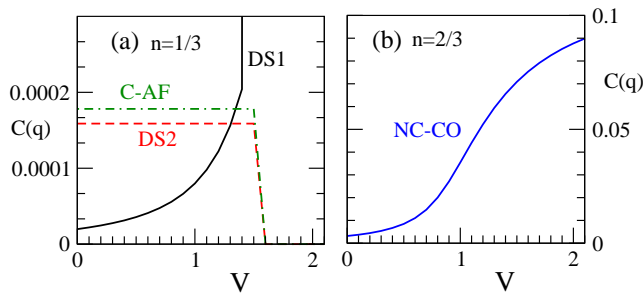}
}
\caption{(Color online) The change in the peak intensity of the charge structure factors at characteristic ${\bf q}$ for the four different spin-charge ordered 
phases at, (a) $n=1/3$ and (b) $n=2/3$, with increasing nn Coulomb repulsion $V$.}
\label{fig4}
\end{figure}

In Fig.~\ref{phase_diag23}, we show the low-T MC energy for different values of $J_{AF}$ at $n=2/3$. Following the analysis at $n=1/3$, we compare the MC energies
with those obtained for ideal ordered spin patterns. We require only three phases to perfectly describe the MC energy data across the 
full $J_{AF}$ range. Two of these phases are the
expected limiting phases: a ferromagnet at small values of $J_{AF}$ and a $120^{\circ}$ YK phase
at large $J_{AF}$. The entire intermediate range belongs to another exotic spin-charge ordered phases. 
A MC snap-shot of this magnetic phase at 
low temperature is shown in inset in Fig.~\ref{phase_diag23}. 
The spin structure remains planar, as in the $120^{\circ}$ phase. In fact, for a specific choice of global orientation, 
all spins are pointing towards the neighboring
sites, which is also similar to the $120^{\circ}$ phase. The important difference is that in this new phase, 
there are three different type of triangles, as shown in the real-space plot in Fig.~\ref{phase_diag23}.
First type is the usual 
$120^{\circ}$ orientation, the second type is formed with two anti parallel spins with the third one pointing at $60^{\circ}$.
The third type of triangle can be obtained from the second type by flipping the spins. 
Similar to other magnetic phases discussed so far,
this magnetic arrangement also generates inequivalent sites in terms of the hopping amplitudes. This inequivalence is reflected via a charge ordering 
pattern (see inset in Fig.~\ref{phase_diag23}) that closely resembles the charge modulations observed in 
various triangular lattice materials \cite{Wawrzynska2008,Wawrzynska2007a,Bernhard2004,Chung2008}.
The electronic DOS in this magnetic phase supports two gaps (see inset in Fig.~\ref{phase_diag23}), 
corresponding to filling fractions of $n=2/3$, and $n=1/3$, thereby justifying the existence of the NC-CO phase at both filling fractions.


In order to test the stability of these unusual spin-charge ordered phases in the presence of electron-electron interactions, we add to the Hamiltonian
Eq. (1) a nn repulsive interaction, $H_1 = V \sum_{\langle ij \rangle} n_i n_j$. An unrestricted Hartree-Fock analysis is performed by keeping 
the magnetic order fixed. The $C({\bf q})$ is then computed for the self-consistent solutions for local charge densities.
We plot in Fig. \ref{fig4} 
the magnitude of the $C({\bf q})$ at characteristic values of $q$ as a function of $V$ for each of the four ordered phases.
Two of the phases at $n=1/3$, DS2 and C-AF, are not affected by the nn repulsive interaction (see Fig. \ref{fig4} (a)).
However, beyond a critical value of $V$, these phases are destabilized in favor of the expected charge ordered phase consisting of a high density
site surrounded by low density sites.
On the contrary, the structure factors for DS1 and NC-CO states increase with increasing $V$, indicating that both these states are further stabilized by a nn 
repulsive interaction.


To conclude, we have reported four new spin-charge ordered 
phases at filling fractions of $n=1/3$ and $n=2/3$ in the strong coupling KLM on a triangular lattice. 
Two of these phases are collinear and stripe-like in their magnetic arrangement, 
and the other two are noncollinear. Presence of magnetically inequivalent sites leads to 
charge ordering in all the phases. The charge ordering pattern for the noncollinear phase at $n=2/3$ is identical to that observed
in various triangular lattice materials with active spin degree of freedom, such as, 2H-AgNiO$_2$, 3R-AgNiO$_2$ and Na$_x$CoO$_2$
\cite{Wawrzynska2008,Wawrzynska2007a,Bernhard2004,Chung2008}. The inclusion of a nn Coulomb interaction enhances the charge ordering further,
indicating that mutually supportive mechanisms could be involved in stabilizing such ordering in real materials.
All the states reported in this work break the discrete rotational symmetry of the triangular lattice via their charge and spin arrangements. 
This may lead to interesting states with charge-spin nematicity at finite temperatures.
The stability of these novel states relies on the nature of
the electronic spectrum for itinerant fermions, which develops a gap at the chemical potential.
Consequently, all the new phases reported in this study are electrically insulating with gap of the order of bare hopping amplitude.
These insulators can neither be called Slater-type nor Mott-type since the 
opening of gap can neither be understood from the Fermi surface nesting arguments nor from the infinite coupling limit.
Therefore, such exotic spin-charge ordered insulators are prototype examples of cooperative many-body effects which are not easy to
understand within effective single particle theories.

We thank M. Daghofer for useful discussions.
SK acknowledges hospitality at IFW Dresden during June-July 2014, and financial support from the Department of Science and Technology, India.

\bibliographystyle{apsrev4-1} 
\bibliography{library}

\end{document}